\documentstyle[11pt,aaspp4,flushrt]{article}

\slugcomment{Submitted to The Astrophysical Journal}

\def\msun{\ifmmode {\rm M_\odot} \else M$_\odot$\fi}
\def\msunyr{\ifmmode {\rm M_\odot~yr^{-1}}\else${\rm M_\odot~yr^{-1}}$\fi}

\def\muobs{\ifmmode {\mu_{o}} \else  $\mu_{o}$ \fi}
\def\mdoto{\ifmmode {\dot{M}_0} \else  $\dot{M}_0$ \fi}
\def\geff{\ifmmode {g_{eff}} \else $g_{eff}$ \fi}
\def\teff{\ifmmode {T_{eff}} \else $T_{eff}$ \fi}
\def\gz{\ifmmode {g_{z}} \else  $g_{z}$ \fi}
\def\grad{\ifmmode {g_{rad}} \else  $g_{rad}$ \fi}
\def\inu{\ifmmode {I_\nu} \else  $I_\nu$ \fi}
\def\jnu{\ifmmode {J_\nu} \else  $J_\nu$ \fi}
\def\knu{\ifmmode {K_\nu} \else  $K_\nu$ \fi}
\def\hnu{\ifmmode {H_\nu} \else  $H_\nu$ \fi}
\def\fnu{\ifmmode {F_\nu} \else  $F_\nu$ \fi}
\def\bnu{\ifmmode {B_\nu} \else  $B_\nu$ \fi}
\def\snu{\ifmmode {S_\nu} \else  $S_\nu$ \fi}
\def\kapf{\ifmmode {\overline{\kappa}_F} \else  $\overline{\kappa}_F$ \fi}
\def\kapnu{\ifmmode {\kappa_\nu} \else  $\kappa_\nu$ \fi}
\def\signu{\ifmmode {\sigma_\nu} \else  $\sigma_\nu$ \fi}
\def\sigt{\ifmmode {\sigma_T} \else $\sigma_T$ \fi}
\def\epsnu{\ifmmode {\epsilon_\nu} \else  $\epsilon_\nu$ \fi}
\def\taunu{\ifmmode {\tau_\nu} \else  $\tau_\nu$ \fi}
\def\cm{\ifmmode {\rm cm} \else  cm \fi}
\def\cmmitwo{\ifmmode \rm cm^{-2} \else $\rm cm^{-2}$\fi}
\def\cmmithree{\ifmmode \rm cm^{-3} \else $\rm cm^{-3}$\fi}
\def\cmps{\ifmmode \rm cm~s^{-1}\else $\rm cm~s^{-1}$\fi}
\def\cmpsps{\ifmmode \rm cm~s^{-2}\else $\rm cm~s^{-2}$\fi}
\def\kmps{\ifmmode \rm km~s^{-1}\else $\rm km~s^{-1}$\fi}
\def\kmpspmpc{\ifmmode \rm km~s^{-1}~Mpc^{-1} \else
    $\rm km~s^{-1}~Mpc^{-1}$\fi}
\def\ergps{\ifmmode \rm erg~s^{-1} \else $\rm erg~s^{-1}$ \fi}
\def\ergpsphz{\ifmmode \rm erg s^{-1} Hz^{-1} \else 
   $\rm erg s^{-1} Hz^{-1}$ \fi}
\def\mdotav{\ifmmode \langle \dot m_o \rangle \else 
   $\langle \dot m_o \rangle$ \fi}
\def\eg{e.g.}

\def\cf{cf.}
\def\etal{~et al.}

\begin{document}

\title{Accretion Disks and the Lyman Continuum Polarization of QSOs}

\author{Gregory A. Shields and Lance Wobus}
\affil{Department of Astronomy, University of Texas, Austin, TX 78712}

\and

\author{Dirk Husfeld}
\affil{Munich University Observatory, Scheinerstr. 1,
D-81679 Munich, Germany}

\lefthead{Shields \etal}
\righthead{Lyman Continuum Polarization of QSOs}

\normalsize

\renewcommand{\theenumi}{\roman{enumi}}
 
\begin{abstract}

HST observations of some QSOs show a strong, abrupt increase
in polarization at rest wavelength about 750 \AA. 
The closeness of the
polarization rise to the H I Lyman edge suggests a
connection, but the displacement to shorter wavelengths,
and the shape of the polarization rise require explanation.

We have computed the polarized spectrum of a
thermally emitting accretion disk around a supermassive
black hole, including the effects of the relativistic
transfer function.  
The local stellar atmosphere spectra show a blueshifted 
polarization rise in the Lyman continuum, as found by
Blaes and Agol (1996).  However, the relativistic transfer
function adds an additional blueshift of sufficient magnitude
that the model cannot explain the observations.

We show that
a good fit results if  the emitted radiation 
is assumed to have a sharp increase in polarized
flux at the Lyman edge in the rest frame of the orbiting gas.
Relativistic effects then cause the observed polarization
to rise sharply at a wavelength substantially less
than 912~\AA.  The blueshift depends on the 
angular momentum
of the black hole and the inclination of the disk.  
A good fit to PG 1630+377 results from a simple model
with a dimensionless angular momentum
$a_* \equiv cJ/M^2 = 0.5$ 
and an observer viewing angle $\mu_o \equiv cos~\theta_o = 0.1$.
Smaller values of $a_*$ give insufficient blueshifts, and
values close to $a_* = 0.9982$ 
require unrealistically large polarizations in the rest
frame of the gas.
An intermediate value of $a_*$
might result from coallescing black holes,
successive accretion events, or electromagnetic extraction of
angular momentum from the hole.

\end{abstract}


\section{Introduction}

The leading model of energy production in QSOs involves an accretion disk
orbiting a supermassive black hole (Blandford 1990).
Evidence for black holes in the nuclei of nearby galaxies is increasing
(Rees 1997), but proof of accretion disks in QSOs has been elusive.

The expected disk effective temperatures imply disk emission at
wavelengths in rough agreement with the Big Blue Bump observed in energy
distributions of QSOs (Shields 1978; Malkan 1983; Czerny and Elvis 1987). 
Accretion disk fits to the energy distributions of a number of QSOs have
been quite successful (\eg, Sun and Malkan 1989, Webb \etal\ 1993).  These
models indicate masses $M_H$ of over $10^9~\msun$ and accretion rates
$\dot{M}$ above 
$1~\msunyr$ for luminous QSOs.  Lyman edges in the total flux should be
inconspicuous because of relativistic Doppler broadening (Laor and Netzer
1989, hereinafter LN; Laor 1991) and NLTE effects (Coleman 1993; Shields and Coleman 1994,
hereinafter SC; Hubeny and Hubeny 1997). The Lyman edge of H I will be most
conspicuous in relatively cool disks ($T_{max} \simeq 25,000~\rm K$),
but will appear as a slope change
over the wavelength range $\sim1100$ to $700$~\AA\ (SC) rather than an abrupt
discontinuity. 

Models of accretion disks predict substantial electron scattering opacity
at optical and ultraviolet wavelengths, leading to polarization of
the emitted light (see section 2.1).  In contrast, observed  polarizations in the optical
and near ultraviolet are
typically $\sim1$ percent or less (Stockman, Moore, and Angel 1984;
Webb \etal\ 1993).  Moreover,
in radio loud objects where a radio extension 
indicates the position angle of the
disk axis, the observed polarization tends to be parallel to the disk
axis (Berriman \etal\ 1990). Electron scattering 
in a geometrically thin disk would produce
polarization perpendicular to the axis.   Predicted polarizations can 
be reduced by absorption opacity, Faraday rotation (Agol and Blaes 1996),
general relativistic depolarization (Laor, Netzer, and Piran 1990, 
hereinafter LNP), and
disk surface structure (Coleman and Shields 1991);
but some concern remains (Antonucci \etal\ 1996).

LNP computed the expected energy distribution and polarization of QSO
disks with a simple model of the atmosphere and full consideration of
the general relativistic transfer function.  They found that while the
Lyman edge in the energy distribution is inconspicuous in general,
a substantial drop in polarization in the H I Lyman continuum should
result from the increase in absorption opacity relative to scattering.
Several groups undertook
spectropolarimetry with the
{\it Hubble Space Telescope (HST)} Faint Object Spectrograph (FOS), targeting
``Lyman edge candidates'' for which International Ultraviolet Explorer (IUE)
spectra suggested an intrinsic drop in the flux around the Lyman edge
(\eg, Koratkar, Kinney, and Bohlin 1992).
Rather than a drop in polarization in the Lyman continuum, these
observations revealed an abrupt rise in polarization at rest
wavelength $\sim 750$~\AA\ in several objects (Impey \etal\ 1995;
Koratkar \etal\ 1995).  The polarization increases by a large factor,
from $\leq 1$ percent at $\lambda \geq 1000$~\AA\ to $\sim 5$ percent
in PG 1222+228 at $\lambda \sim 600$~\AA\ 
and $\sim 20$ percent in PG
1630+377. The proximity of the polarization rise to the Lyman edge is
suggestive (Impey \etal\ 1995).

An intriguing explanation of the displaced polarization rise was offered
by Blaes and Agol (1996, hereinafter BA).  Using LTE and NLTE models for the
disk atmosphere with a hydrostatic vertical structure, they found that,
for $T_{eff} \approx 20,000$ K, a similarly blueshifted, steep
polarization rise resulted from a combination of effects.
The polarization drops at the Lyman edge because of the increased absorption
opacity, but then it increases with decreasing wavelength because of the
$\lambda^{-3}$ dependence of the H I photoionization opacity and
the increasingly steep source function gradient at higher frequencies.
A strong source function gradient leads to severe limb darkening,
needed for strong polarization (Agol, Blaes, and Ionescu-Zanetti 1997; Cheng
\etal\ 1988). A model for a disk around a nonrotating black hole gave
a promising fit to PG 1222+228, but polarization
as strong as observed for PG1630+377 could not be achieved.  However,
BA did not include the relativistic transfer function, and it was
unclear whether these effects would upset the fit to PG1222+228.

We report here results of a theoretical study of polarization
in QSO accretion disks. We focus on the steep polarization
rises observed in PG 1222+228 and PG 1630+377. In Section 2, we describe our
stellar atmosphere models and the treatment of the relativistic transfer
function.  Then we describe attempts to fit the two QSOs
with accretion disk atmospheres, and show that general relativistic
effects preclude an acceptable fit.  In Section 3, we show that
a good fit can be achieved in a simple model with a sharp polarization jump
at the Lyman edge in the rest frame of the gas, and we consider
possible sources for this emission.  Our conclusions are discussed
in Section 4.

\section{Accretion Disk Model Atmospheres}

\subsection{Computational Methods}
We have calculated LTE stellar atmospheres for conditions appropriate for
QSO disks.  These models were supplemented by some ``mixed" LTE-NLTE
calculations as described below.  Because we are interested only in
the continuum, and because a perfect
description of the vertical structure of AGN disks is not available,
we used simple methods to model the atmosphere.  These
resemble the LTE calculations of Coleman (1993), but the codes
used here were written largely independently. The physics included is
similar to the treatment of BA, but the computational methods are quite
different. 

Atmospheres were computed for a set of radii and corresponding values of
effective temperature,
\teff, and vertical acceleration of gravity, \gz, evaluated for a
given $M_H$, $\dot{M}$, and angular momentum, $a_* = cJ/GM^2$, from the
relativistic relations given by Novikov and Thorne (1973) and Page and Thorne
(1974).
The atmosphere was computed for a set of points (typically 91) equally
spaced in the logarithm of the column density  (gm cm$^{-2}$) ranging from
$\rm log~\Sigma(z) = -6.5$ to 2.5. The
hydrostatic density profile was integrated from the top down according to 
the equation
%
$$dP_g/d\Sigma = g_{eff} = g_z - g_{rad},$$
%
where $P_g$ is the gas pressure and
$g_{rad} = \kapf F/c$ is the radiative acceleration.
Here 
$$\kapf \equiv \int_{0}^{\infty} F_\nu \kappa_\nu d\nu
/\int_{0}^{\infty} F_\nu d\nu$$
is the flux-weighted mean opacity per
unit mass (Mihalas 1978).

Radiative transfer in the atmosphere was calculated by performing a downward
integration of the equations 
%
$$d\knu/d\taunu = \hnu$$
%
and
%
$$d\hnu/d\taunu = \epsnu(\jnu-\bnu),$$
%
where \jnu, \hnu, and \knu are the usual Eddington moments of the radiation
field $\inu(\taunu, \mu)$, \taunu is the monochromatic optical depth from
the top of the atmosphere, $\epsnu \equiv \kapnu/(\kapnu+\signu)$, and
\kapnu and \signu are the absorption and scattering opacities per unit
mass.  The emergent flux \hnu was adjusted in a shooting method
to satisfy a lower boundary condition \jnu = \bnu at a
fiducial optical depth, typically
$\taunu = 8$.  The first iteration used a gray atmosphere temperature profile,
the Eddington approximation (\jnu = 3\knu), and
a surface boundary condition $\jnu = \sqrt{3} \hnu$. Subsequent iterations
used variable Eddington factors and a temperature profile corrected by
the Uns\"old-Lucy method  to achieve flux conservation (Mihalas 1978).
We assumed constant flux, i.e., no local heating (see below).
The resulting \jnu(\taunu) was used
to compute the source function \snu(\taunu), and new values of \jnu,
\knu, and \hnu were then
computed at each depth point from the usual exponential integral expressions
(Mihalas 1978).  These gave new values for the Eddington factors, 
the  quantities needed
for the temperature correction, and \kapf. The entire
procedure (hydrostatic integration, transfer integration, moments,
temperature correction) was iterated as necessary, typically three iterations.
  Opacity sources included were electron scattering, 
free-free absorption,
and bound-free absorption from the lowest 7 levels of H I and the
ground states of He I and He II.  Ionization and H I level populations
were  computed from the Boltzmann-Saha
equation (LTE). Convergence was good except in a thin surface layer
for certain ranges of
\teff, for which H or He were partially ionized at the surface.  The
flux in the related Lyman continuum of H or He was negligibly small in these
cases, and the fluxes at longer wavelengths were not significantly affected.

The assumption of constant flux through the atmosphere 
should be adequate
in our case
(see also BA).  We describe below models using the same parameters
($M_H, \dot{M}, \alpha$) as BA. The region of the disk emitting the Lyman continuum
is characterized by the radius $9.5GM/c^2$ giving maximum \teff (25,500 K).
  In the vertically averaged model of NT with $\alpha = 0.1$,
the surface density of the disk (midplane to surface) is $\Sigma_c
= 2000~\rm g~cm^{-2}$.  The surface density above the point where
\taunu = 1 at $\lambda = 911$ \AA\ is $\Sigma_1 = 0.8~\rm g~cm^{-2}$.
If energy is dissipated along the vertical extent of the disk roughly
in proportional to density, then only a small fraction of the flux is
dissipated in the photosphere.  Also, we find that, at \taunu = 1, the dissipation
rate per unit mass is tiny compared with the radiative heating and cooling rates
in the atmosphere.  These
conclusions are even stronger if dissipation is proportional to gas pressure or
total pressure. Moreover,  our conclusions depend
more on relativisitic effects
than atmospheric details.

Polarization was calculated using a straightforward iterative procedure
by Voigt (1951) applied to the final converged atmosphere.  Although
Voigt studied cases with small scattering opacity fractions,
we found that convergence
was rapid in all cases.  The exact
polarization exceeds that of the first iteration by almost a factor two for
wavelengths at which electron scattering strongly dominated the opacity.  This
results from multiply scattered photons that produce substantial
``prepolarization'' of the radiation going into the last scattering. 

CS and others have emphasized that the effective gravity \geff in
AGN disks will be very low because of close cancellation of \gz
and \grad.  This leads to a large gas pressure scaleheight, a
low density in the photosphere, high ionization,
and a large opacity contribution from electron scattering.  Coleman
(1993), Hubeny and Hubeny (1997), and others have shown 
that significant differences
in emitted spectrum can result from use of a fully self-consistent
vertical structure rather than an approximate, constant gravity
solution.  However, our results show that, for the polarization
phenomena of primary concern here, the details of the local atmospheric
emission are less important than the effect of the relativistic
transfer function.  Therefore, we followed the procedure of BA in
using a value of \gz that was 20 percent larger 
than the Eddington limit for a given
\teff. This is about the minimum \gz that allowed convergence
for most \teff.  
For these purposes, we took $g_{Edd}
= \kappa_{es}\pi F/c$, where $\kappa_{es}$ is the electron scattering
opacity for fully ionized pure hydrogen.  
(Reduction of \gz to $1.1g_{Edd}$, in
cases where the atmosphere still converged, gave considerably
lower values of \geff, but the emergent spectrum and polarization
changed only modestly.)

Sensitivity of our results to NLTE effects was evaluated in several
sample cases by using
our LTE atmospheric structure, $\rho(\Sigma)$ and $T(\Sigma)$, as input to
an NLTE stellar atmosphere program developed by the Munich
University Observatory stellar atmosphere group.
This code, using modern methods including accelerated lambda
iteration (ALI), includes bound-bound and bound-free transitions of
many levels of H and He.  When set to run in LTE mode, this code
gave emergent spectra in excellent agreement with our LTE code.
In NLTE mode, the results showed the expected reduction of the Lyman
absorption edge for relatively cool atmospheres
(\cf~Coleman 1993; Hubeny and Hubeny 1997).  However, the polarization
as a function of wavelength was not greatly different between the
LTE and NLTE cases, in agreement with BA.

Following BA, we cut off the disk at $50 R_g$, where $R_g \equiv
GM/c^2$ is the gravitational radius.  For Schwarzschild black holes,
we computed atmospheres at $r_* \equiv R/R_g$ = 7.76, 9.52, 15.3, 20.3, 31,
44, and 50.  Test cases with additional radii gave similar
results.  Summation over disk
radii and application of the relativistic transfer function, 
which accounts for
Doppler shifts, transverse and gravitational redshifts,
and gravitational deflection of light rays was accomplished with
a modified version of the computer codes used by LNP and kindly made available
by T. Piran and A. Laor.  These codes use a set of 51 radii
from $r_*$ = 1.2346 to 400 to cover disks around
either rotating or nonrotating black holes.  We typically used
as set of 80 frequencies uniformly spaced in log $\nu$ between
$1.96\times10^{14}$ to $1.85\times10^{16}$ Hz.

The LNP transfer
function assumes limb darkening laws in the intensity and
polarization for pure electron
scattering (Chandrasekhar 1960), although there is provision for a
scale factor in the polarization.  The actual
limb darkening in the polarized intensity can be quite different (\cf\
Cheng \etal\ 1988; Agol, Blaes, and Ionescu-Zanetti 1997; and references
therein). In order to accomodate a general limb darkening law with minimum
changes to the LNP codes, we used a simple fitting procedure.  The polarized
intensity from a given radius and frequency can be described in terms
of the orthogonal polarized intensity components $I_r(\mu)$ and $I_l(\mu)$.
Each of these was fit with a six term series in integer powers
of $1-\mu$ from n = 0 to 5, which gave an excellent fit in all cases. 
Heuristically, this may be thought of as a set of 12 accretion
disks, each of which emits at its surface 
either pure $I_l$ or pure $I_r$, with
a limb darkening $(1-\mu)^n$.  The corresponding 12 transfer functions
were computed with the LNP code and applied to the respective fitting
components, and the
resulting Stokes parameters at the observer were summed to produce the observed
spectrum. Our atmosphere results were interpolated to the LNP radius grid
and the adopted frequency grid
using a linear interpolation in $R$ and $\nu$  for (1) the brightness
temperature
$T_\nu$ defined by $F_\nu = \bnu(T_\nu)$ and (2) $I_r(\mu)/I_r(0)$ 
and $I_l(\mu)/I_l(0)$
at five values of $\mu$ (prior to determining the fitting coefficients). 
Test cases gave good agreement with a disk emitting with the
electron scattering darkening law and with results from an
independent transfer function code by Agol(1997a,b).

The emergent energy distribution and polarization
for an atmosphere with $\teff = 25,500$ and ${\rm log}~g_z = 
{\rm log}(1.2g_{Edd})
= 2.58$ are shown in Figure 1. The polarization
rises in the Lyman edges of H I and He I are evident, 
similar in the case of H I
to the results of BA (who did not include He).
Note also the substantial polarization at optical wavelengths (see also
Figure 3 of BA).  The low effective gravity leads to a low
density at the optical photosphere, and electron scattering contributes
substantially to the opacity ($\sim 75$ percent at $\tau_\nu = 1$ for
$\lambda = 5000$~\AA).  Some earlier models (\eg, LNP) used the midplane
density throughout the atmosphere.  The predicted optical polarizations
may be a problem for disk models, given the weak observed optical
polarizations for radio quiet QSOs (\eg, Berriman \etal\ 1990).

\subsection{Results}

We have computed models for PG 1222+228 and PG 1630+377, for which the
most dramatic Lyman continuum polarization rises have been reported.
For PG 1222+228, following BA and Webb \etal\ (1993), 
we have used a nonrotating black hole with 
$M_9 \equiv M_H/10^9\ \msun = 5.3$ and
$\mdoto \equiv \dot{M}/1\ \msunyr = 18.9$. 
(Actually, only the quantity $\dot{M}/M_H^2$ matters for the shape of
the energy distribution. This, along with $a_*$, determines the run of \teff
with radius, and $g_z$ is determined by \teff in our models.  
The overall luminosity can be scaled to fit the observations by
scaling $\dot{M}$.  The same is true for the ``toy'' models
of Section 3.)
The resulting energy distribution is shown in  Figure 2 for
observer viewing angle $\muobs \equiv \rm cos(\theta_{o}) = 0.35$.
(Larger $\muobs$ give weaker polarization; the same model for
$\muobs$ = 0.1 is shown in Figure 3.)
The energy distribution is in fairly good agreement with the observations,
as found by BA and Webb \etal\ (1993).  However, the polarization
rise is more blueshifted and more gradual than observed.  This differs
from the results of BA, who did not include the relativistic transfer
function (but who warned that it could modify their conclusions).
Note that for \muobs = 0.35, more edge on than BA's model,
the polarization still reaches the observed $\sim 5$ percent, but at
a wavelength shorter than observed.  The transfer function diminishes the
peak polarization and adds its own blueshift to that already present in the
stellar atmosphere spectra, causing the total blueshift
to exceed that
observed.   

Figure 3 shows the spectrum of the same disk, viewed at a more edge
on angle (\muobs = 0.1) and scaled to the observed energy distribution
of PG 1630+377.  The energy distribution fits fairly well.
However, even for this
nearly edge on viewing angle, the polarization does not approach the value
$\sim 20$ percent observed, and the polarization rise is again too much
blueshifted in the model.

Another difficulty with the BA model may be the small ionizing luminosity
from such a cool disk.  The expected H$\beta$ equivalent width for our
model is only $\sim 25$ \AA\ for \muobs = 0.35 if the line emiting gas absorbs
all the ionizing photons.  Observed equivalents widths are somewhat larger,
and the covering factor for the emitting gas may be small.  Moreover,
the disk's He I and He II ionizing luminosities are negligible.  However,
a ``nonthermal'' ionizing source may also be present.

{\em We conclude that the stellar atmosphere phenomenon suggested
by BA to explain the observed polarization rises fails on account
of the relativistic transfer function.}  The actual stellar atmosphere effect
is nevertheless present, and it may contribute at some level to the
observed spectrum of AGN disks.

\section{A Polarization Jump ``At the Edge''}

The preceeding results show that the relativistic transfer
function causes a substantial blueshift of any polarization
rise in the disk's local emission.  This raises the question, what will be
observed if the disk gas emits a continuum with an abrupt increase in
polarization right at the Lyman edge in the rest frame of the gas?  We have
calculated a simple model in which the disk surface, at any radius, emits a
continuum with a sharp Lyman edge in absorption in the total flux together with
a sharp jump in polarization at the Lyman edge.  The emitted astrophysical
flux was given by $\fnu = 4\hnu = \bnu(\teff)$ for $\nu < \nu_H$
and $\fnu = \bnu(0.841\teff)$ for $\nu \geq \nu_H$, simulating an atmosphere
whose brightness temperature drops to the boundary temperature in the
Lyman continuum.  The polarization was assumed to rise from effectively zero
at $\nu < \nu_H$ to an arbitrary 
multiple of $p_{es}(\mu)$ at $\nu \ge \nu_H$, where
$p_{es}$ refers to the polarization of electron-scattering
atmosphere (Chandrasekhar 1960).
To this end, we define $a_p$ by $p(\mu) = a_p p_{es}(\mu)$.

Figure 4 shows the resulting fit to PG 1630+377 for several values of
the black hole angular momentum and \muobs = 0.1.  
For each value of $a_*$, the
values of $M_9$ and $\mdoto$ were adjusted to fit the energy distribution,
and $a_p$ was adjusted to fit the maximum polarization observed.
Figure 4 shows that the observed polarization
rise remains quite sharp and is displaced to the blue by an amount that
depends on $a_*$.  For larger $a_*$, the disk's inner boundary at the
marginally stable radius shifts
inward, from $r_*$ = 6 for $a_0 = 0$ to $r_*$ = 1.22 for $a_*$ = 0.9982
(Page and Thorne 1974).  For larger $a_*$, the Lyman continuum comes
from smaller radii where the relativistic effects are stronger.
The wavelength and shape of the polarization rise agree with the
observations for $a_*$ = 0.5.
Evidently, {\em this simple model gives a good fit to the observed
energy distribution and to the shape of the rise in polarization.}

Why does the polarization remain low for a considerable wavelength
interval to the blue of the Lyman edge, and then rise abruptly? The
relativistic transfer function causes the approaching, blueshifted part of the
disk to appear brighter than the receding portion.  In addition, light from the
sideways moving portion of the disk behind the black hole is bent
as it passses near the hole on its way to the observer.  
Even for a fairly edge-on
viewing angle, this light leaves the atmosphere fairly close to 
the normal, so that its intensity is enhanced by limb darkening, 
the projected solid angle
subtended by unit surface area is increased, and its polarization
leaving the disk surface is low (\cf~Agol 1997a).  
This light also suffers depolarization from relativistic rotation
of the polarization plane, which varies rapidly with position on the
part of the disk behind the black hole 
(Connors, Piran, and Stark 1980; Agol 1997a).
The light from the part of the disk in front of the hole is
more strongly polarized but less intense.
Thus, the observed light at small Doppler shifts is
dominated by the back part of the disk and is weakly polarized.
At larger blueshifts, the light is dominated by the approaching
part of the disk.  This light is strongly polarized because it 
suffers less gravitational deflection and therefore emerges
from the atmosphere at a large angle to the normal. These points
are illustrated in Figure 5, which shows the received flux and polarization
for an annulus  that emits at a single radius, $r_*$ = 10.52,
and wavelength, $\lambda = 912$ \AA.  The received emission
from the near and far sides of the ring are shown separately.  The
flux is much stronger at the blue end than at the red end of the
observed range of wavelength shifts because of Doppler boosting.  
Intense but weakly
polarized flux is received from the back side of the disk with relatively
little shift from the emitted wavelength.  Thus, the Lyman edge in the total
flux will ramp up or down progessively from the red to the blue side
of the nominal wavelength, whereas the polarization will rise only at
shorter wavelengths. 

The blueshift of the polarization rise in the toy model
depends on \muobs as well as $a_*$.  Figure 6 shows results for
a maximally rotating hole with $a_*$ = 0.9982.  Again, $\dot{M}$
and $M_H$ are adjusted in each case to fit the energy distribution.
The blueshift of the polarization
rise increases with decreasing \muobs, and by eye we estimate 
a best fit \muobs $\approx$ 0.33. However, in this case, large values of the
intrinsic polarization, $a_p \approx 19$, are  required to match the
observed {\em degree} of polarization of PG 1630+377.  This corresponds 
to $p(\mu) = 76$ percent at $\mu$ = 0.33 and 
$p(\mu) > 100$ percent at $\mu < 0.2$.  The unknown mechanism for
producing the postulated Lyman continuum polarization could have a limb
darkening law quite different from $p_{es}(\mu)$, but polarization
of roughly 76 percent at $\mu \approx 0.33$ will likely be required
in any case and may be difficult to
achieve.  If one rejects such a large intrinsic polarization, then
$a_*$ must be less than 0.9982 in the context of this toy model.

The cases \muobs = 0.25 and 0.50 are shown in Figures 7 and 8.
For \muobs = 0.25, the blueshift is best fit for $a_* \approx 0.7$.
For \muobs = 0.5, the blueshift depends little on $a_*$ (even for
$a_* = 0.9982$; see Figure 6), and it is too small to fit PG 1630+377.
Figures 4 and 6--8 show that the wavelength of the polarization rise provides
a constraint in the ($a_*$, $\mu$) plane. Interpolating, we find
\muobs = (0.10, 0.25, 0.30) for $a_*$ = (0.50, 0.75, 0.90).
From the fit to the energy distribution, we find \mdoto = (27, 25, 20),
$M_9$ = (5.0, 9.0, 15), and $L/L_E$ = (0.22, 0.14, 0.099),
respectively.  Note that $\dot{M}$ decreases with increasing $a_*$ because 
of increasing efficiency of energy production.  Also, $\dot{M}/M_H^2$
decreases with increasing $a_*$ because of stronger Doppler boosting
(requiring lower \teff) and a smaller radiating area in units of
$(GM/c^2)^2$, which gives higher \teff for a given $\dot{M}/M_H^2$.  
All of the values are comfortably below the
Eddington limit, consistent with a thin disk.

This toy model was also used to fit PG 1222+228.   
The blueshift of the
observed polarization rise is somewhat less than for
PG 1630+377 (Impey \etal\ 1995; Koratkar \etal\ 1995),
and the rise possibly is more abrupt, but the uncertainties
are large.  Figure 9 shows results for \muobs = 0.25.  The
fit for $a_*$ = 0.5 is fairly good, requiring $a_p$ = 1.7, \mdoto 
= 43, $M_9$ = 6.2, and $L/L_E = 0.26$.  Reference to Figure 7
suggests that \muobs = 0.5 would fit PG 1222+228 
fairly well for appropriate $a_p$,
independent of $a_*$, although the blueshift may be slightly
less than observed.

Figures 6, 7, and 8 indicate that the wavelength at which the polarization rises
to one-half of its full value, $\lambda_{1/2}$, will be between about 640 and 830 \AA,
at least for inclinations \muobs $\leq 0.5$, which are most likely to give strong
polarization.
This is consisstent with the two objects discussed above and with
PG 1338+416 (Koratkar \etal\ 1995).

The observed degree of polarization for the toy model scales with
$a_p$, and results can be derived from the figures.  For example,
if we fix $a_* = 0.5$ and $a_p = 4.4$ (above discussion of
PG 1630+377), we may ask what will be the plateau
value of the polarization, $p_{max}$, for this disk viewed at various
angles.  From Figures 4, 7, and 8, we find $p_{max} = (20, 13, 5.0)$ percent
and $\lambda_{1/2} = (750, 770, 810)$ \AA\ for $\muobs = (0.1,0.25, 0.5)$.

\section{Discussion}

We have shown that the relativistic transfer function precludes
an explanation of
the observed polarization rises in PG 1222+228
and PG 1630+377 in terms of the stellar atmosphere effect
proposed by Blaes and Agol (1996).  Relativistic effects add an additional
blueshift to the polarization rise that makes the predicted rise
occur at wavelengths shorter than observed.  The model also cannot
explain the strong degree of polarization observed in PG 1630+377.

As a step toward a more successful model, we have shown that
the observed polarization rises are consistent with a simple model
in which the polarization rises abruptly
at $\lambda912$ in the rest frame of the orbiting gas.
The observed polarization rise is blueshifted
by an amount that depends on $a_*$ and \muobs. 
For PG 1630+377, the observations are consistent with
$a_*$ of about 0.5 or greater but probably less than 0.9982.

What could be the origin of the polarized Lyman emission?  In
PG 1630+377, the polarized flux, $\fnu_p = p_\nu \fnu$ rises steeply
with decreasing wavelength (Koratkar \etal\ 1995).  An increase in
polarized flux abruptly at the Lyman edge in the rest frame of the
gas suggests free-bound emission from ionized hydrogen that undergoes
electron scattering, either in the emitting gas or a separate
scattering location (\cf\ Impey \etal\ 1995; Koratkar \etal\ 1995).  
In view of the evidence of irradiation of
AGN disks by a hard incident continuum, at least in lower luminosity
objects (\eg, Clavel \etal\ 1992;
Mushotzky, Done, and Pounds 1993; 
and references therein), one naturally
thinks of a photoionized layer on the disk surface
(\eg, Sincell and Krolik 1997).
A simple picture would be a uniform slab of ionized gas in which
recombination of H I produces a fairly uniform emissivity
in Lyman continuum photons that scatter off electrons in the 
slab on their way out of the slab.  Such a surface layer, overlying
a photosphere with a Lyman edge in absorption in its total flux,
might resemble our toy model.  Results by Phillips and
M\'esz\'aros (1986) show polarizations of 15 percent
or more for $\mu \approx 0.1$ in a scattering slab with uniformly
distributed emitters, and the possibility of even larger polarizations
for emitters concentrated near the surface. 
The strongest polarization occurs for $\tau_{es} \approx 0.2$ to 0.3. 
However, it is not clear from their results that 
strong enough polarization
for PG 1630+377 will be possible, and the limb darkening law is 
very different from $p_{es}(\mu)$.

The temperature of such
a slab can be constrained by the fact that the polarized flux in PG 1630+377
jumps by an order of magnitude across the polarization rise.  If
the polarized flux has the same energy distribution as the radiation
feeding into the scattering process, then the emissivity of the
gas must have a similar contrast.  The emissivity ratio 
$j_\nu(\lambda 912-)/j_\nu(\lambda 912+)$ decreases with increasing temperature
because of free-free and Balmer continuum emission.
The observed polarized flux requires $T \leq 80,000$ K for PG 1630+377,
based on expressions by Brown and Mathews (1970).  

A layer that is mechanically heated rather than photionized may be
an alternative source of the polarized Lyman continuum. 
Yet another possible source involves the fact
that for luminous QSOs, a standard (Shakura and Sunyaev 1973) disk may
become optically thin in the inner radial zone.

Further work is needed to establish
whether the continuum and line emission in such a model is consistent
with all known observations for QSOs with Lyman polarization rises.
Koratkar \etal\ find a polarization rise at the expected position
of Lyman $\alpha$, weaker than the Lyman 
continuum polarization and not blueshifted.
This may come from larger disk radii with lower orbital velocities.
Simple estimates, based on the black body limit for the L$\alpha$ surface
brightness, suggest that the disk radii considered here will not be a
significant source of L$\alpha$ emission.

The spin rates of black holes in AGN have received increasing attention in
recent years.  An interesting ``spin paradigm'', in which radio loud objects
have rapidly rotating holes (Blandford 1990) has been developed by Wilson and
Colbert (1995).  Values $a_* \approx 0.5$ could arise in several ways.
(1)  If two black holes coallesce, the resulting hole has
$a_* \approx 0.5$ for a mass ratio $\ge 0.2$ if roughly half the angular
momentum of the binary at the last stable circular orbit is radiated as
gravitational waves during the coallescence (Wilson and Colbert 1995).
(2)  Successive accretion events at random inclinations could cause the hole's
angular momentum
to wander around a value $a_* \approx 0.5$ if the mass accreted in one
event is not too small as a fraction of the black hole mass (Moderski,
Sikora, and Lasota 1997).
(3)  Angular momentum extraction by the Blandford-Znajek (1977)
mechanism may balance angular momentum supplied by accretion to give an
equilibrium value $a_{*,eq}$.  Results by Moderski and Sikora (1996)
indicate $a_{*,eq} \approx 0.5$ for $\alpha \dot{m} \approx 0.01$,
where $\alpha$ is the viscosity parameter (Shakura and Sunyaev 1973)
and $\dot{m} \equiv c^2\dot{M}/L_{Edd}$.  Our model for PG 1630+377 has
$\dot{m} \approx 1$, and values
$\alpha \approx 10^{-2}$ are not precluded in AGN.  However, a value
of $a_*$ as large as 0.5 might correspond to an object with substantial
radio luminosity in the ``spin paradigm'', whereas PG 1630+377 and PG 1222+228
are radio quiet objects.

The toy model discussed here essentially addresses the question, what must
the polarization be in the rest frame of the gas, in order for the observed 
polarization rise to match the observations?  The remarkably 
simple answer is that it
should be an abrupt rise in polarization and polarized flux at the Lyman
edge.  This may provide a significant clue to the source of the polarized
radiation.

\section{Acknowledgments}

We thank E. Agol, R. Antonucci, O. Blaes, 
E. Fierce, A. Koratkar, R. Kudritzki, M.
Malkan, and the referee for helpful discussions and comments, 
and H. Coleman for the use of several
numerical routines. This material is based in part upon work
supported by the Texas Advanced Research Program under Grant No. 
003658-015 and by the Space Telescope Science Institute under
Grant No. GO-06044.03-94A..

\clearpage

\centerline{{\bf Captions for Figures}}


\figcaption[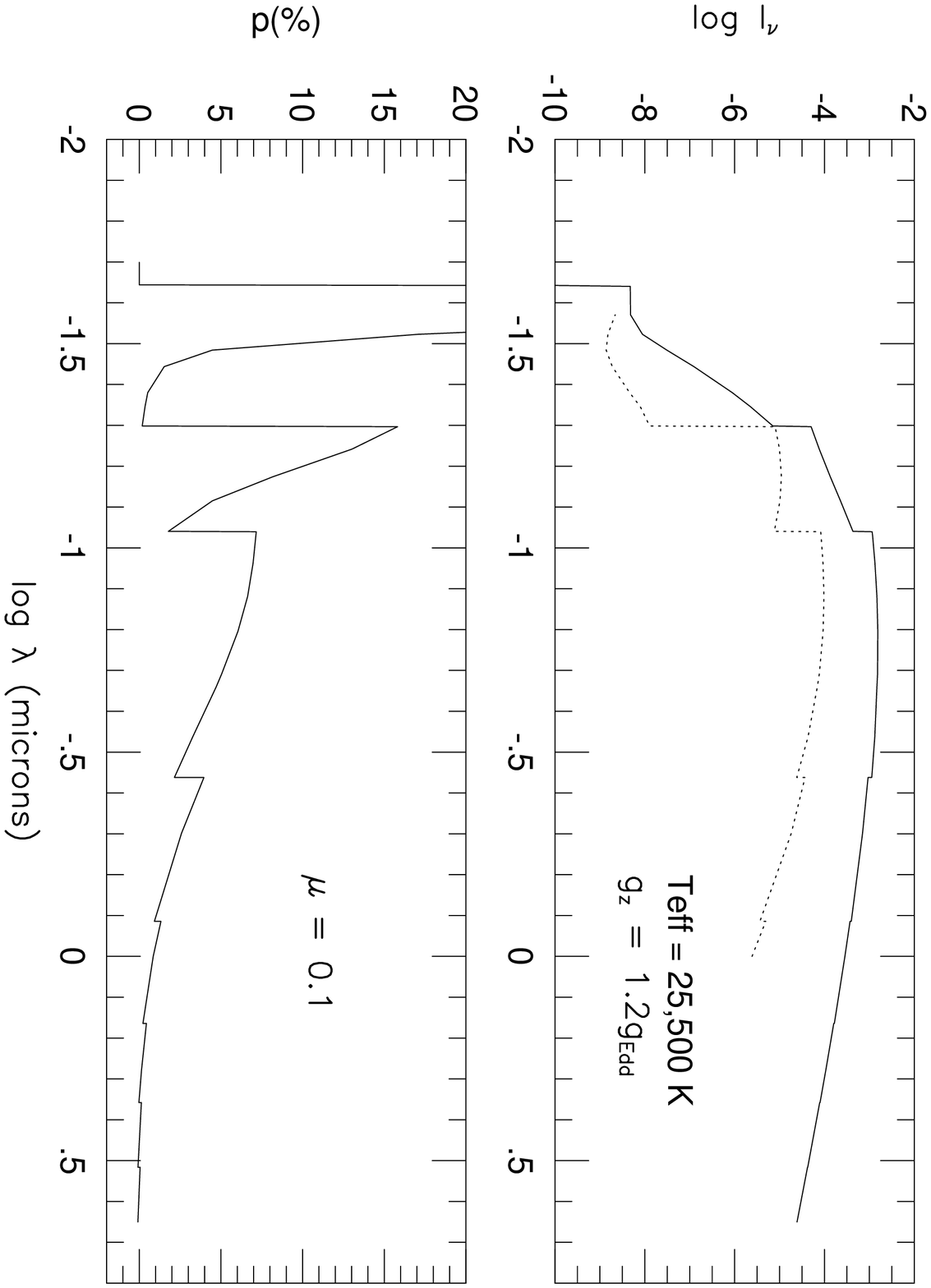]{Emergent intensity, polarized intensity (dashed), 
and polarization
at angle $\mu = 0.1$ from an LTE atmosphere with \teff = 25,500 K
and $g_z = 1.2g_{Edd} = 10^{2.58}~\cmpsps$.  
This \teff corresponds
to a radius $r_* = 9.5$, where \teff is highest, in a disk with
$m_9$ = 5.3 and \mdoto = 18.9, used to model PG 1222+228.  \label{fig1}}

\figcaption[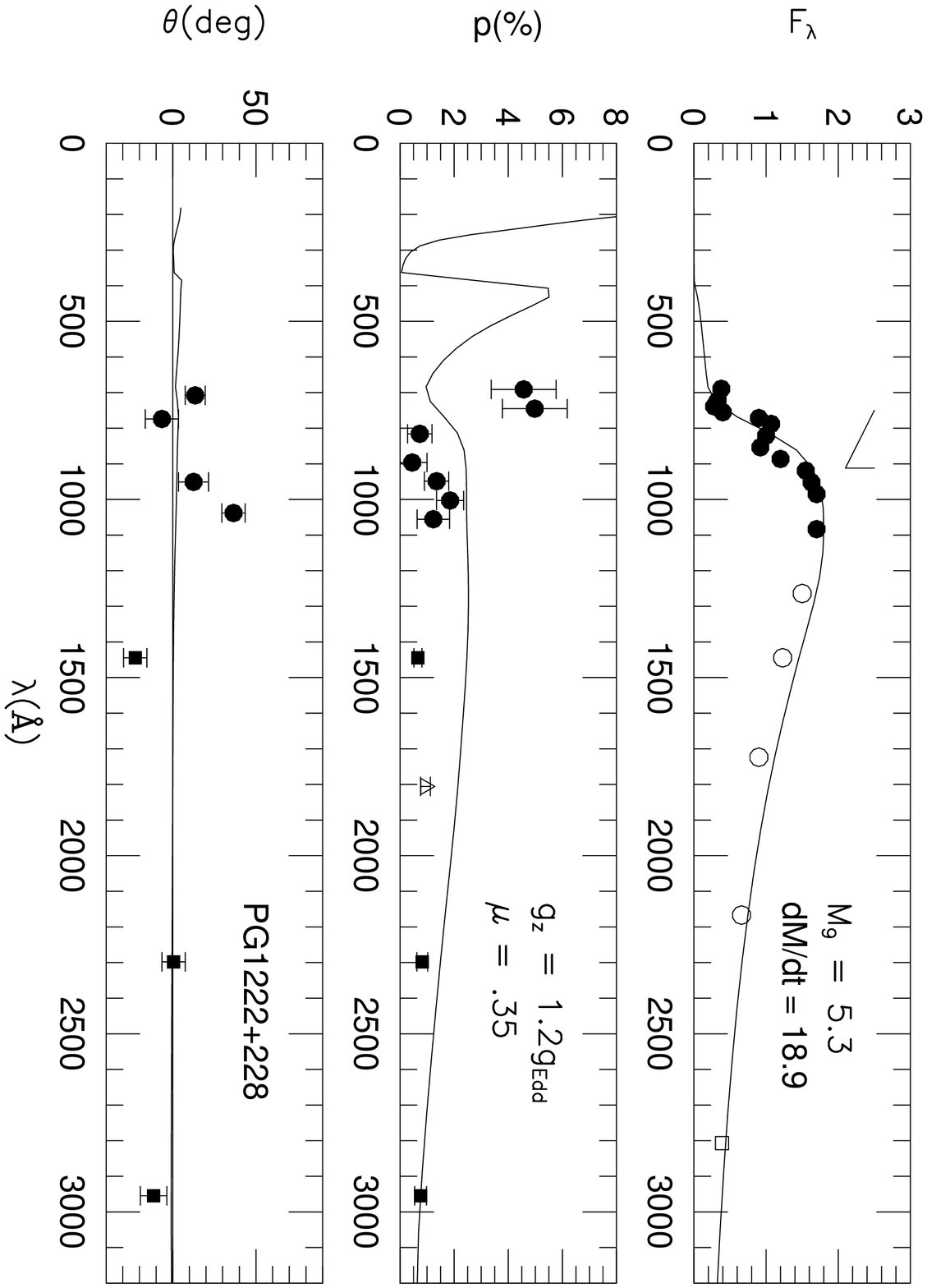]{Accretion disk model compared with 
observations of PG 1222+228.
Observations are from Impey \etal\ 1995 (filled circles), Bechtold \etal\
1984 (open circles), Wampler \& Ponz 1985 (open square), 
Webb \etal\ 1993 (filled squares), and Stockman \etal\ 1984 (open triangles).
Model disk with
$M_9$ = 5.3 and \mdoto = 18.9 is viewed at angle \muobs = 0.35. 
Position angle of the disk axis is unknown, so there is an arbitrary
offset between model and observed position angles.
Abscissa is rest wavelength.  
The position of the Lyman edge redshift is
indicated in the top panel.
$F_\lambda$ has been scaled to fit the observations,
corresponding to modest adjustments in $M_H$ and $\dot{M}$
at fixed $\dot{M}/M_H^2$ (see text). \label{fig2}}

\figcaption[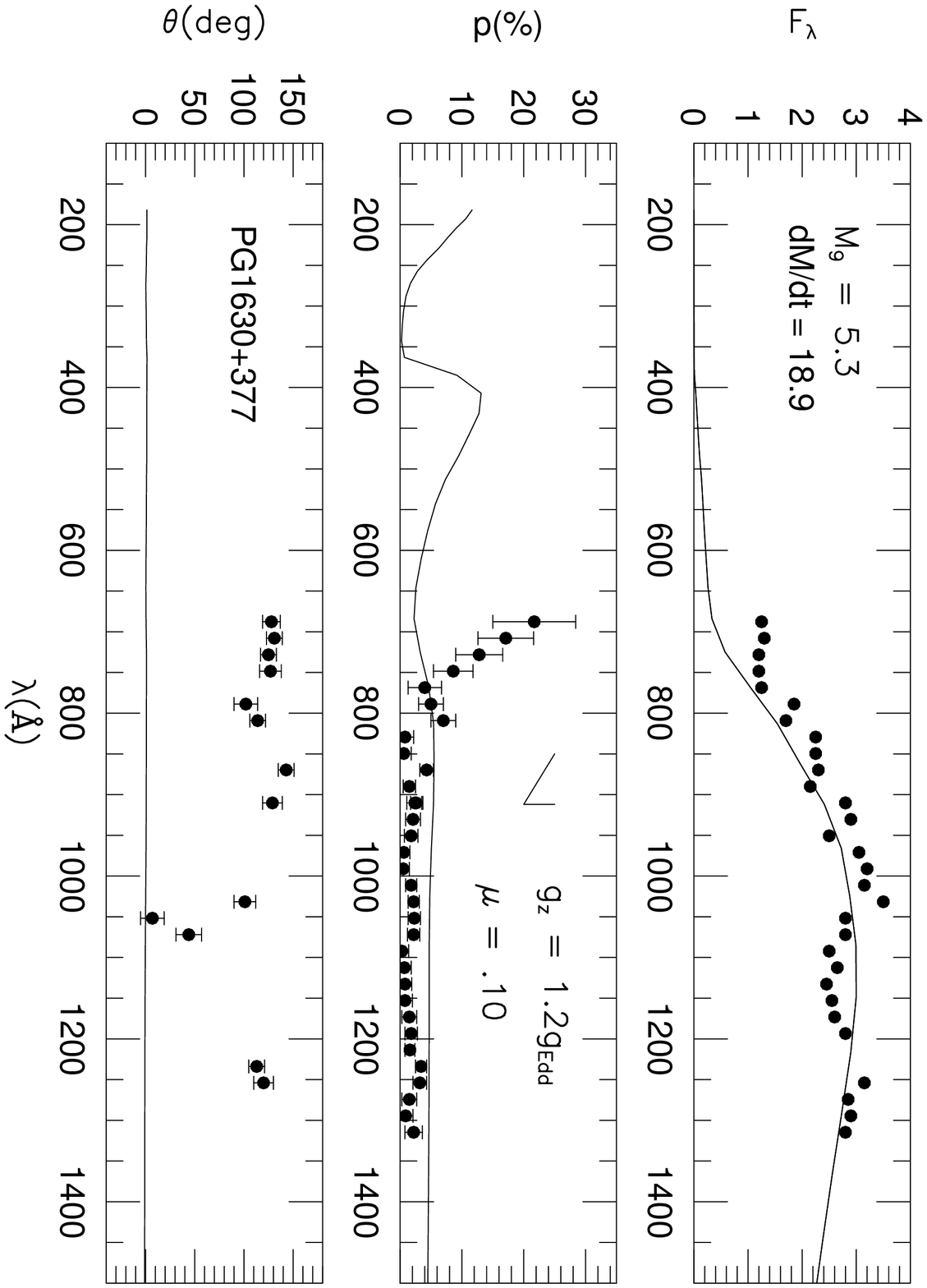]{ Accretion disk model compared with observations of
PG 1630+377.  Observations are from Koratkar \etal (1995).
Model has been scaled at constant $\dot{M}/M^2$ 
from that for PG 1222+377 so as to fit
the observed maximum flux, leaving $\teff(r_*)$ unchanged.  
Viewing angle is \muobs = 0.1.
Position angle of the disk axis is unknown, so there is an arbitrary
offset between model and observed position angles. \label{fig3}}

\figcaption[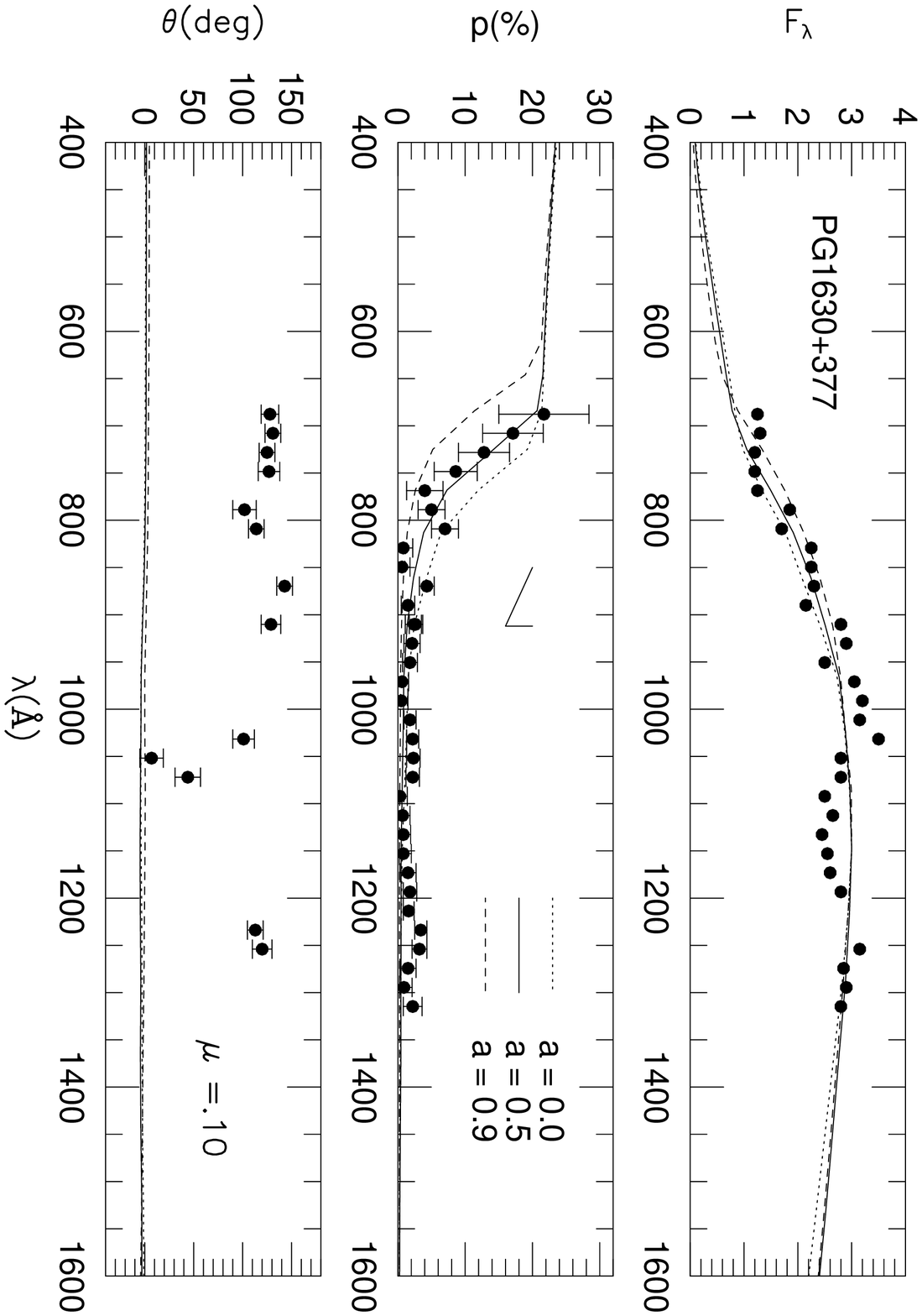]{Toy accretion disk model compared with observations of
PG 1630+377 (see text).  Fluxes have been scaled to fit maximum
observed $F_\lambda$.  Viewing angle is \muobs = 0.1.  For
each of the three values of black hole angular momentum, $a_*$, 
the value of $\dot{M}/M^2$ has been adjusted to fit the 
observed energy distribution. The wavelength of the polarization
rise is increasingly blueshifted for larger $a_*$.
The models have $p/p_{es}$ =  (4.4, 4.4, 5.7) for 
$a_*$ = (0.0, 0.5, 0.9). \label{fig4}}

\figcaption[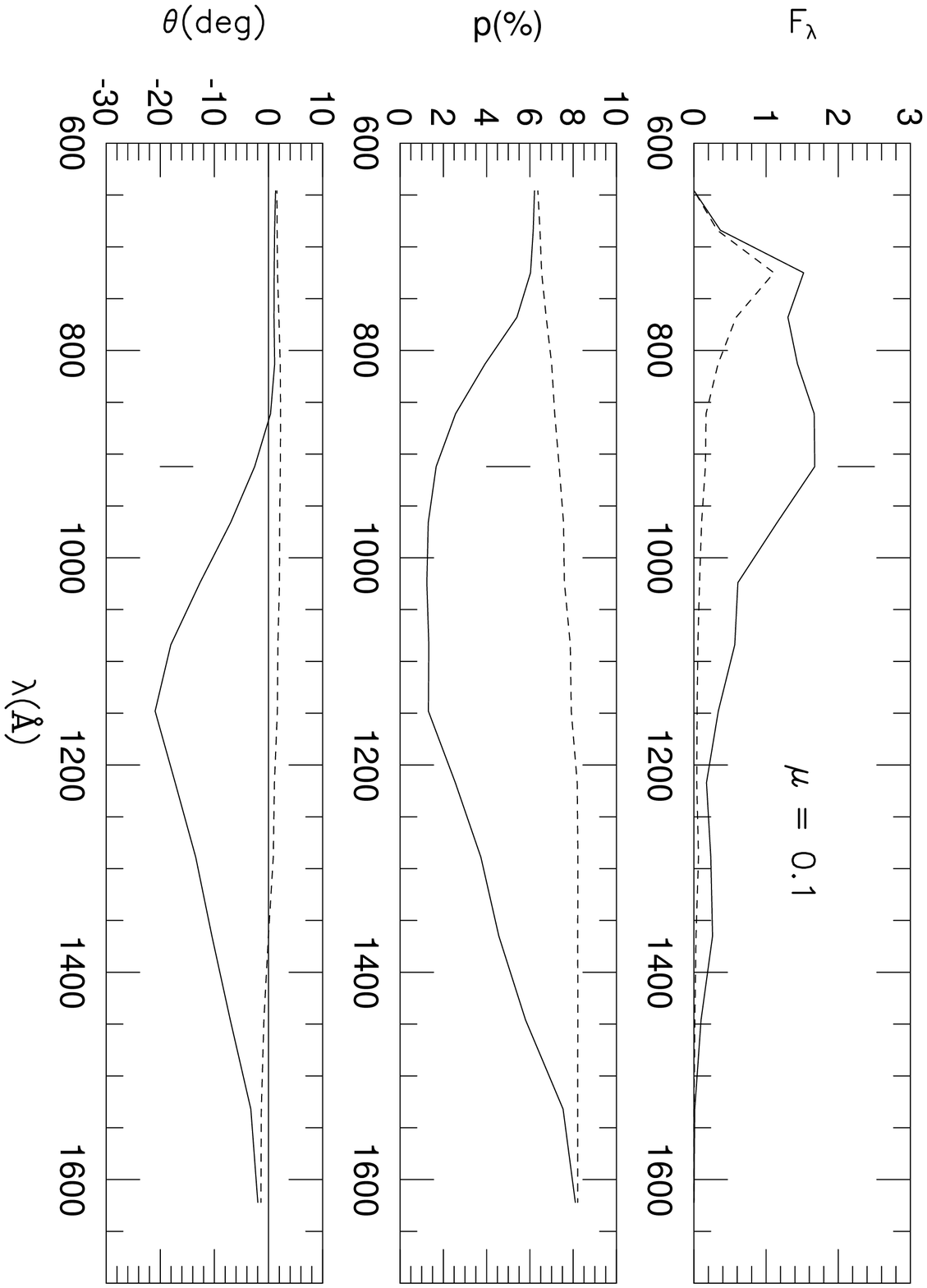]{Predicted flux, polarization, and positional angle
for a narrow ring orbiting a black hole with $a_* = 0.5$
at a radius $r_* = 10.52$.
The observer viewing angle is \muobs = 0.1.
The emitted spectrum is a delta-function 
at $\lambda 912$, with the 
intensity and polarization in the rest frame of the gas 
depending on angle in the manner of
a pure electron scattering atmosphere (Chandrasekhar 1960.)
The light received from the far side and the near side 
of the disk are shown separately as the solid and 
dashed curves, respectively. \label{fig5}}

\figcaption[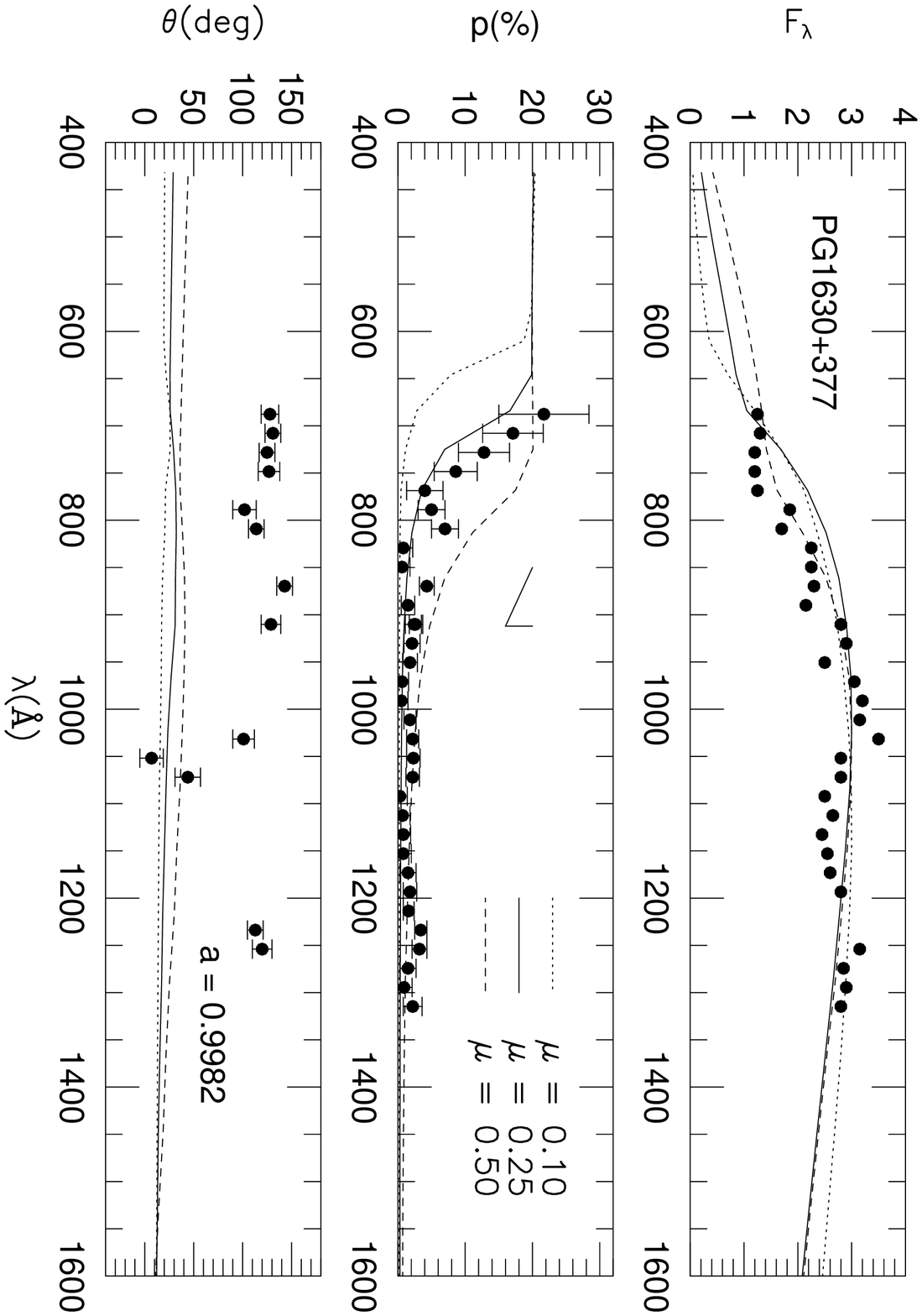]{Toy model results for $a_*$ = 0.9982 for several viewing
angles, compared with observations of PG 1630+377.
The models have $p/p_{es}$ =  (9.0, 14.9, 30.8) for 
\muobs = (0.1, 0.25, 0.5). \label{fig6}}

\figcaption[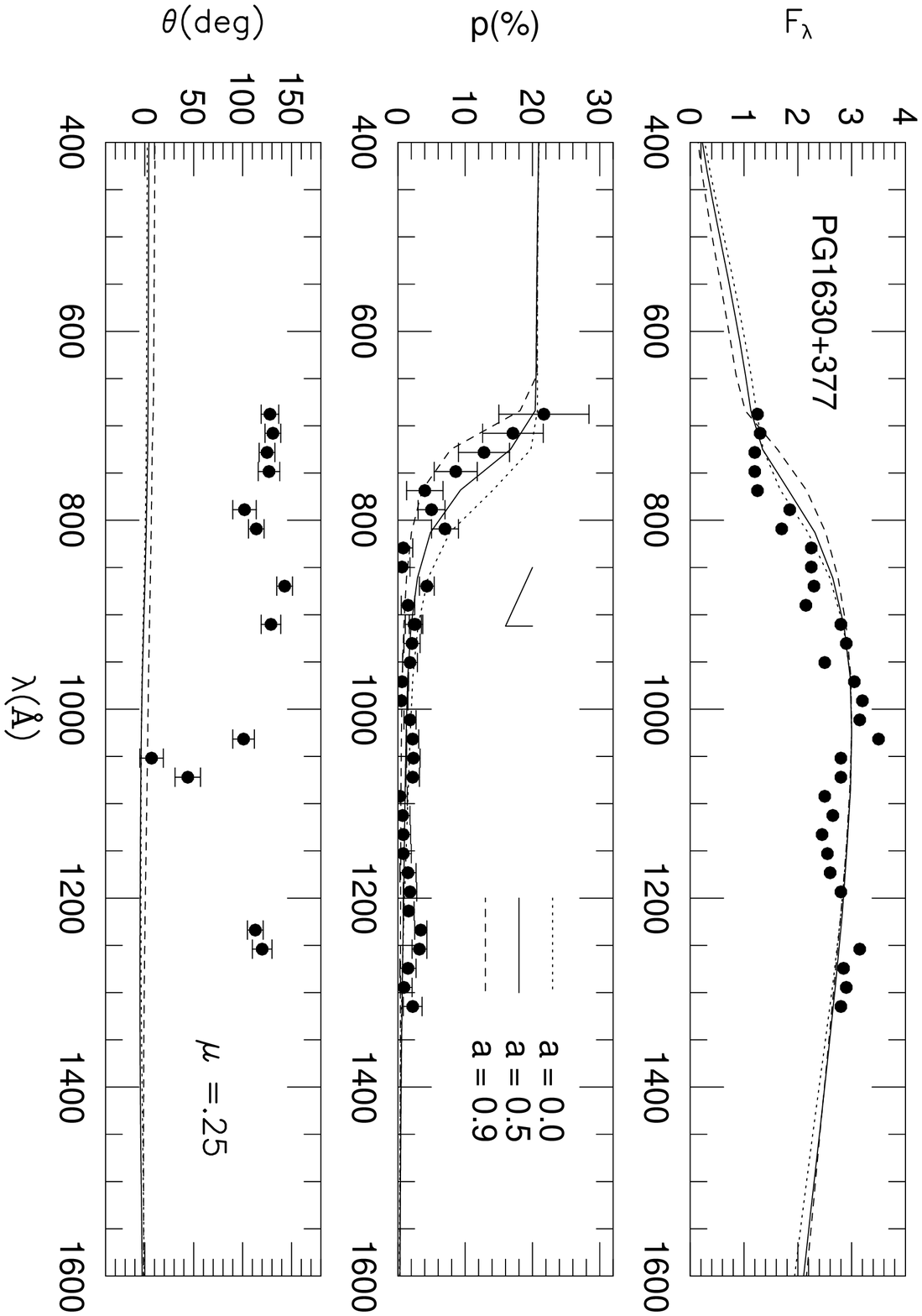]{Toy model results for \muobs = 0.25 and several values
of $a_*$, compared with observations of PG 1630+377.
The models have $p/p_{es}$ =  (6.4, 7.1, 9.6) for 
$a_*$ = (0.0, 0.5, 0.9). \label{fig7}}

\figcaption[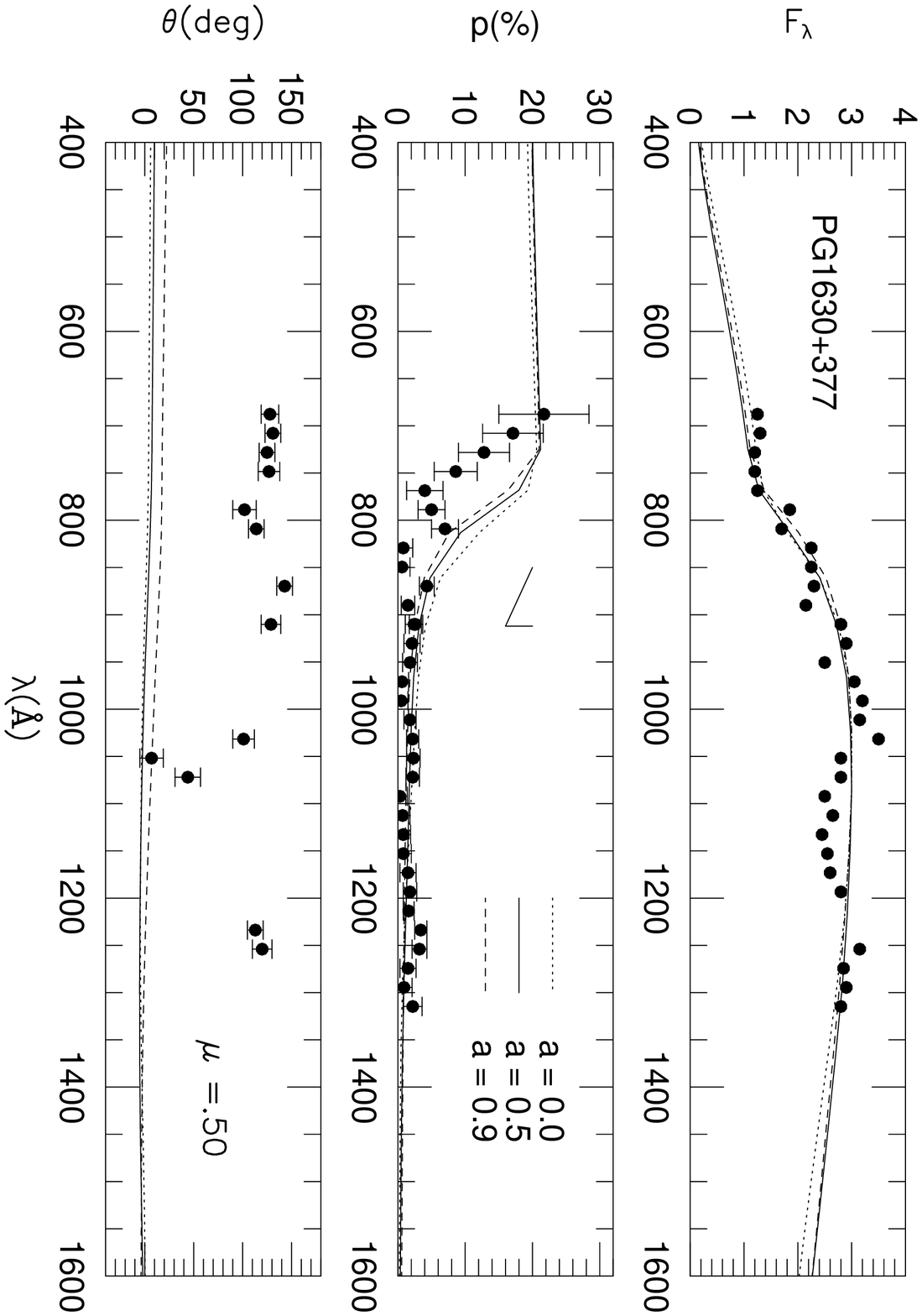]{Toy model results for \muobs = 0.5 and several values
of $a_*$, compared with observations of PG 1630+377.
The models have $p/p_{es}$ =  (14.0, 17.5, 24.0) for 
$a_*$ = (0.0, 0.5, 0.9). \label{fig8}}

\figcaption[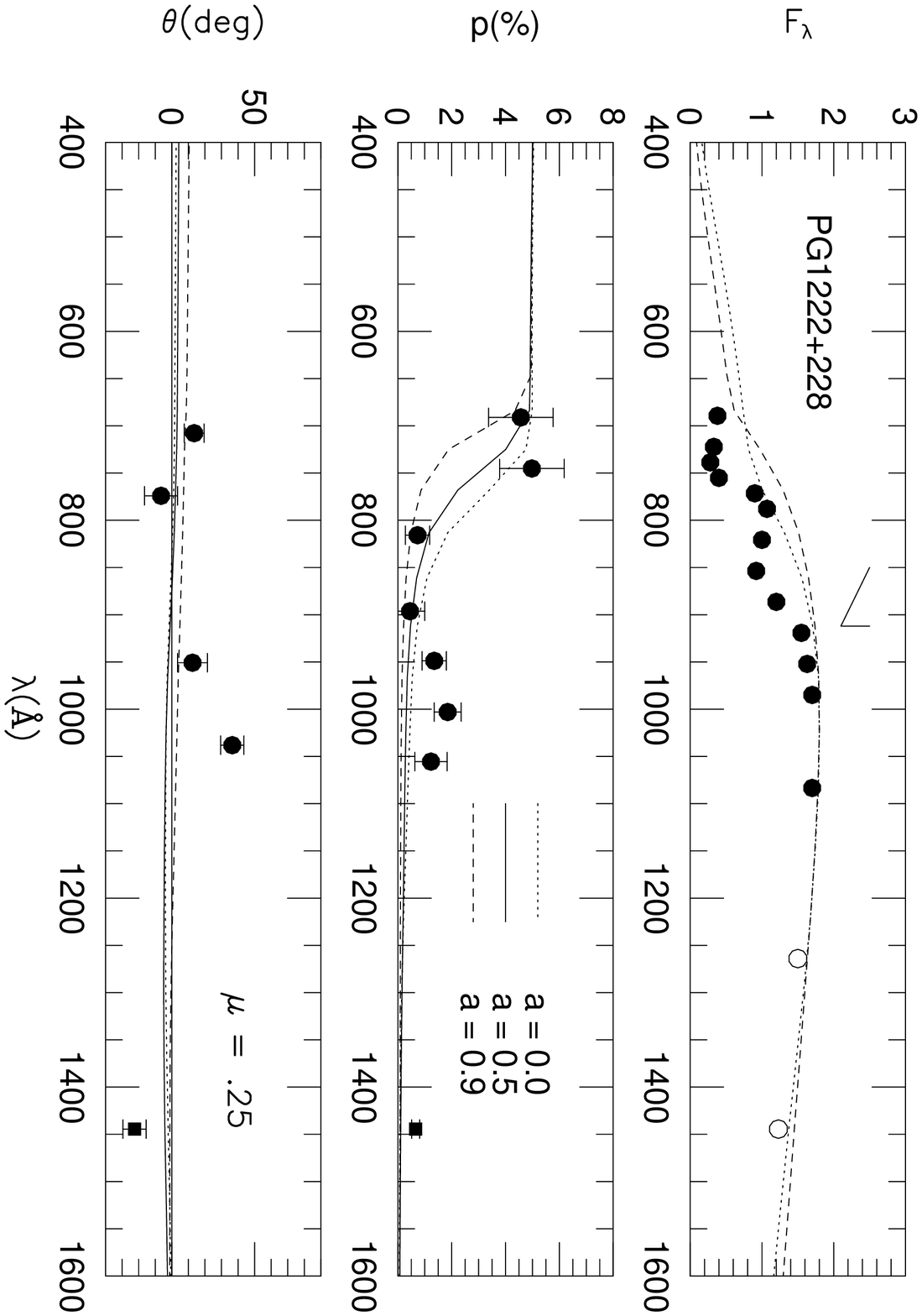]{Toy model results for \muobs = 0.25 and several values
of $a_*$, compared with observations of PG 1222+228.
Values of $a_p$ are one fourth as large as for the corresponding
models applied to PG 1630+377 (Figure 7). \label{fig9}}

\clearpage

\plotone{fig1.ps}
\clearpage
\plotone{fig2.ps}
\clearpage
\plotone{fig3.ps}
\clearpage
\plotone{fig4.ps}
\clearpage
\plotone{fig5.ps}
\clearpage
\plotone{fig6.ps}
\clearpage
\plotone{fig7.ps}
\clearpage
\plotone{fig8.ps}
\clearpage
\plotone{fig9.ps}
  
\end{document}